# Evaluation of Enhanced Security Solutions in 802.11-Based Networks

Ajah Ifeyinwa Angela

Department of Computer Science, Ebonyi State University Abakaliki, Nigeria

**ABSTRACT**

*Traditionally, 802.11-based networks that relied on wired equivalent protocol (WEP) were especially vulnerable to packet sniffing. Today, wireless networks are more prolific, and the monitoring devices used to find them are mobile and easy to access. Securing wireless networks can be difficult because these networks consist of radio transmitters and receivers, and anybody can listen, capture data and attempt to compromise it. In recent years, a range of technologies and mechanisms have helped makes networking more secure. This paper holistically evaluated various enhanced protocols proposed to solve WEP related authentication, confidentiality and integrity problems. It discovered that strength of each solution depends on how well the encryption, authentication and integrity techniques work. The work suggested using a Defence-in-Depth Strategy and integration of biometric solution in 802.11i. Comprehensive in-depth comparative analysis of each of the security mechanisms is driven by review of related work in WLAN security solutions.*

## 1. INTRODUCTION

Wireless network has gained wide deployment due to numerous benefits such as user mobility, rapid and cheap installation, flexibility, scalability, and increased productivity it offers. In addition, rapid advances in this technology with improved capabilities which is seen in third generation (3G) and fourth generation (4G) wireless devices make it attractive for enterprise to run their business. However, the use of this novel technology does not go without security risk. 802.11 network also referred to as WLAN is challenged by lack of physical protection in the medium. Moreover, the fact that WLAN device are ship with all security features disabled make it a playground for hackers to tread on. Higher percentage of these hacker use to to access internet freely and others use it for malicious activites. Traditional WLAN that relies on WEP has security flaws that were revealed in FMS attack (2001), Korek attack (2004), PTW attack (2007), and ChopChop attack (2008), [1]. The consequences of unsecured WLAN are very dangerous to users and business enterprise. Attacks pepertrated on the networks have adverse effects on both individual users and business enterprises. Currently, the availability of more secured security solutions and security deployment best practises has greatly addressed security issues of the legacy protocols. This has made many wireless deployments to be probably more secure than the wired LAN. This is possible because many of the IEE802.11 standards (802.1X, 802.11i) offer security features designed to resolve long standing weakness or address newly discovered ones. Researchers have proved the standards to be vulnerable and these are cited in section two of this work. To the best of my knowledge this is the first worked that holistically surveyed and evaluated the outcome of these past researches on security protocols; 802.1X, WPA , WPA2 ,in juxtaposition to VPN SSL, and VPN IPsec. The objectives of the study are; to give the reader a solid grounding in enhanced WLAN security concepts and technologies solutions at a glance. It will equally allow the network administrator to fully assess the risks associated with using wireless and how to mitigate the risks. The main significance of this work is to inform and encourage WLAN users to adopt security measures and best practices that help bring down their risks to a manageable level. The discussion is focused on their operations, performance impact, limitations, possible attacks on them and recommendation of biometric integration on the solution and best practices for addressing threats to WLAN. This work did not discuss WEP but provided good references for accessing details in WEP. The performance metrics used is based on Strength of algorithm used in encryption, authentication and integrity, and are weighted on a four point (weak, strong, fairly strong and very strong). In addition, susceptibility to attacks, and cost of implementation are also indicated. The key highlight of the work is that there is no one solution that guarantees highly secured network. WLAN requires implementing security at all wireless security layers which include wireless signal security, connection security, data protection, device security, network protection, and end user protection. The work is presented in the following order: Reviewed literature on WLAN attacks and solutions is done in section2. Section 3 gives an overview of enhanced WLAN security solutions. Section 4 presents the comparison of WLAN security Protocols. This document presumes that readers have understanding of WLAN architecture, possible attacks on them and flaws in WEP and the details on the aforementioned background knowledge can be accessed in [2][3]. Readers are encouraged to tailor the solutions and recommended guidelines to meet their specific security and business requirements.

## 2. LITERATURE REVIEW

The IEEE 802.11 standards and several researchers have made significant contributions on WLAN security aspects. It was examined in [4] how users from the general public understand and deal with privacy threats associated with Wi-Fi use. It was found that users lack knowledge of immediate risks, and this made them unmindful of privacy and security in using Wifi. It was recommended that privacy and Wifi security problem can be effectively handled through end-user awareness tools and improving wifi infrastructure. An approach for providing an end-to-end wireless security for local area network client/server environment was proposed and implemented in [5]. Performance of encryption algorithms— Rivest Cipher 4 (RC4) and Advanced Encryption Standard (AES) in terms of time, memory and power were evaluated for different devices, key sizes, and the cryptographic algorithms with and without transmission. The result of the work showed that both RC4 and AES performed similarly in just about all cases. But when transmission of data is considered, RC4 performed slightly better than AES in lightweight devices such as on pocket PC that have very limited processing power and memory. The results question the reality of relying on secret key crypto system in establishing credentials and data protection. A research by [6] on IEEE 802.11-2007 Media Access Control (MAC) security in union to IPsec concludes that the security provided by the 802.11 standard is successful in defending against many popular attacks including: session hijacking, denial-of-service attacks against the authenticator, man-in-the-middle attacks, forgery attacks, data manipulation attacks, fragmentation attacks, iterative guessing attacks, redirection attacks, and impersonation attacks but denial-of service attacks against the supplicant are still possible to achieve. The paper concludes that implementing Internet Protocol Security (IPsec) concepts in any part of the Robust Security Network Association (RSNA) would successfully prevent denial-of-service attacks. IPSec is aimed at Denial of service (DoS) attacks, Spoofing attacks and Man-in-the-middle attacks (MITMs). It was report in [7] that Verisign has issued two digital certificates to someone who improperly posed as a representative of Microsoft. This indicates that a sufficiently motivated attacker's ability to forge credentials may exceed the ability of the Public Key Infrastructure (PKI) provider to detect them. There is need to adopt in IPSec, a mechanism that permits secure administration and distribution of keys such that public keys are associated to intended entities that owns them. WPA and IEEE 802.11i was evaluated in [8] to be robust to a lot of attacks. However, they presented scenarios that produce a DoS attack and DoS flooding attacks on these protocols and finally proposed Static and Dynamic 4-WayHandshake solutions to avoid the attacks. Secure Angle enabled access point that uses angles of arrival (AoA) was designed to operate alongside existing wireless security protocols to enable a "virtual fence" that drops frames injected into the network from a client physically located outside a building, and prevent malicious parties from spoofing the link-layer address of legitimate clients. In preventing spoofing attack, the administrator manually certifies a legitimate client's signature and compares it against the incoming packets. They envisaged a challenge of signatures changing to some degree when obstacles in the environment move, and therefore are of the opinion that signatures must be tracked and updated. However, the exact signature specifications are left as an open research question [9]. In my own view, it appears that the Secure Angle will work on only 802.11n APs as there is no statement on whether other IEEE 802.11 specifications such as 802.11a, 802.11b, and 802.11g can support the system. Also I perceived that this approach does not support user mobility which is one of the benefit of WLAN. Address Resolution Protocol (ARP) [10] spoofing was demonstrated on WPA-enabled network and the work equally showed that in a real world scenario, using strong password in WPA network is brekable [1]. It was argued in [11] that WPA was not really cracked. According to him, the vulnerability seen in WPA was due to a poor deployment of WPA-PSK. He stated that a simple 10-character alpha-numeric random PSK (or greater) will make WPA unable to be cracked with dictionary attacks. The benefits of Wi-Fi Protected Access 2 (WPA2) over previous protocols and the vulnerabilities was explored in [12]. Available modes to secure a wireless network using the WPA2 protocol were also discussed. Suggestions on how the protocol vulnerabilities might be addressed through enhancements or new protocols were made. Patch solution and a new algorithm for dummy authentication key-establishment was proposed in [13]. Study conducted by [14] demonstrated weakness in nonce construction and transmission mechanism used in Counter Mode with Cipher Block Chaining Message Authentication Code Protocol (CCMP) that made it vulnerable to time memory trade off (TMTO) pre-computation attack. The solution on how the attack can be prevented was not given. A biometrics approach to WLAN security was presented in [15], in which user's iris was used by the network server for authentication.

## 3. OVERVIEW OF WLAN SECURITY EXTENSIONS

Service Set Identifier (SSID), MAC Address Filtering and wired equivalent protocol (WEP) were the original 802.11 specifications by the IEEE for securing WLAN. These mechanisms lead to a number of practical attacks that demonstrate their failure to achieving security goals. Alternative security mechanisms such as 802.Ix, 802.11i, SSL, and IPsec, were created to enhance WLAN security and are discussed below. Virtual Private Networks (VPN) is deployed in an enterprise wired network for secured data transmission. Its success made it attractive to wireless developers and administrators as security option in an enterprise WLANs

VPN technology provides three levels of security; Authentication, Encryption and Data authentication. VPN server authenticates every user that uses VPN client to connect to the WLAN. It provide data confidentiality by encrypting the traffic passing through the secure tunnel created on top of the in-secured internet medium. It also guarantees that all traffic is from authenticated devices. The network topology diagram in Fig. 1 presents a hybrid solution where both VPN and Wi-Fi security are deployed in an enterprise network.

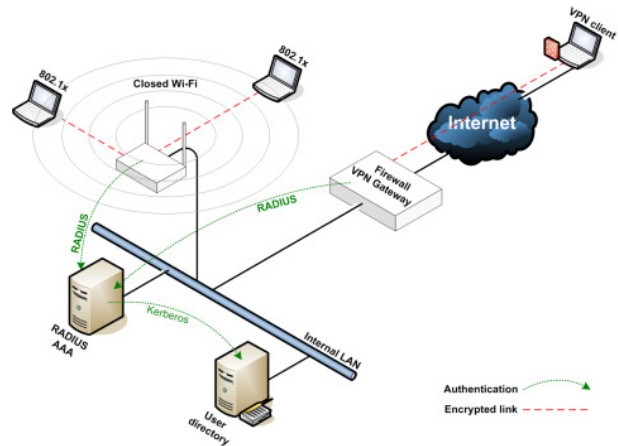

Figure 1. VPN and Wi-Fi security

The VPN gateway permits secured access for internet users and the access points offers secured connectivity for local devices. Wi-Fi association is granted through access control and authentication. Encryption occurs at layer two and above and is hardware based. Centralized RADIUS authentication model is implemented here, in which the network access devices (access points and VPN gateway ) forward RADIUS authentication requests to the RADIUS server for verification. This is used to provide true single sign-on for both Wi-Fi and VPN security.

## 3.1. IPsec

IPsec is "a framework of open standards for ensuring secure private communications over IP networks through the use of cryptographic security services." It consists of two separate security protocols, Authentication Header (AH) and Encapsulated Security Protocol (ESP) that ensures the authenticity and integrity of the data. AH authenticates packets by signing them. The signature is specific to the packet being transmitted, and therefore prevents the data from being modified (integrity).  In addition to handling the authenticity and integrity of data ESP ensures data confidentiality through encryption, digital signature, and/or secure hashes. When AH and ESP are implemented together the entire packet is authenticated. IPSec uses Internet Key Exchange (IKE) mechanism to authenticate end users and manage secret keys by providing a secure exchange of a pre-shared key before IPSec transmissions begin. IKE is not reliable and could end up in a dead state. Dead-Peer-Detection was an improvement in IKE to handle dead state occurrence due to reliability flaws. IPSec uses security association (SA) to describe how parties will use AH and encapsulating security payload to communicate. The SA can be established through manual intervention or by using the Internet Security Association and Key Management Protocol (ISAKMP). ISAKMP approach has advantage over IKE; duplication of functionality in each security protocol is minimal and it uses little time in setting up Communication. IPSec is a widely used VPN protocol.  IPSec VPNs can be deployed in either transport mode or tunnel mode. In transport mode the IPSec-protected data is carried in IP packets that use the original IP addresses of the two VPN peers. Transmission here is faster since the IP headers are not encrypted, and the packets are smaller. The disadvantage in this mode is that a hacker can sniff the network and gather information about end parties. The Tunnel mode encapsulates and encrypts all IP packets and ensures end-to-end transmission using new IP header of the two VPN peers. Tunnel mode is used in host-to gateway or gateway-to-gateway VPNs. IPsec uses Triple DES (3DES), or Advanced Encryption Standard (AES) to ensure confidentiality of IP traffic.

## 3.2. Secure Socket Layer (SSL)

SSL consists of two protocols, the SSL record protocol that defines the format used to transmit data and the SSL handshake protocol that uses the record protocol to exchange messages between the SSL-enabled server and the client when they establish a connection [7]. SSL is used in VPN to protect transmitted data. SSL VPN solution for WLAN provides the following functionality; Centralized security and management, strong and scalable data encryption for maximum security and to secure most sensitive transactions such as online banking on the Internet, "auto reconnect." feature of SSL VPN supports mobility of users, scalability and provides endpoint security. SSL addresses the need of confidentiality integrity and authentication. Confidentiality is achieved by using public key cryptography. Data integrity is preserved by performing a special calculation (hash function) on the contents of the message and storing the result with the message itself. The SSL protocol uses Message authentication codes (MAC) to provide data integrity. It uses Ron Rivest,Adi Shamir, and Leonard Adleman (RSA) algorithm for encryption. The result of the algorithm depends both on the message and the key used. The attacker has no access to the key, and therefore lacks capacity to modify both the message and the digest. SSL uses digital certificates for authentication. Digital certificates contains information such data as the server's name, public encryption key, and the trusted Certificate Authority (CA).

Connection between the client and server is negotiated through a handshake procedure. Here, the client connects to the SSL-enabled server and requests that the server sends back information in the form of a digital certificate. The server can require the client to present a valid certificate as well but this is optional. The client checks the validity of the server certificate. When the handshake phase ends, data exchanged between the client and server is then used with hashing functions to generate session keys that are used for encryption, decryption and tamper detection of data throughout the SSL session. According to [16], quantum cryptography that relies on photon and uses key longer than 128-bit or 256-bit keys will provide better protection than RSA algorithm. It will make data communication potentially unhackable and guarantee security against future computer improvements.

## 3.3. 802.IX/EAP

IEEE 802.IX is specified for port-based network access control for wired networks and has been extended for use in wireless networks. It provides user-based authentication, access control and implements Extensible Authentication Protocol (EAP), which was originally designed for Point-to Point Protocol (PPP). EAP is a layer 3 protocol and can use any authentication mechanism to add flexibility to 802.1X . 802.IX uses;

1. **The Supplicant:** A station that requests access to the network offered by the authenticator. It dialogue with the authentication server through the authenticator.
2. **The Authenticator:** Typically a wireless access point that controls the state of each port (open/close) and mediates an authentication session between the client and the authentication server.
3. **The Authentication Server:** This is a Remote Authentication Dials In User Service (RADIUS) server that do the authentication process for the authenticato.

802.1X and EAP approach are characterized by three main elements that make it better than the basic 802.11 securities. They are:

1. Mutual authentication between client and authentication (Remote Access Dial-In user Service (Radius) server.

2. Encryption keys dynamically derived after authentication
3. Centralized policy control, that stimulate re-authentication and fresh encryption key generation when a session expires.

EAP Authentication Process is shown in Fig.2 and the sequence of events are numbered 1-9.

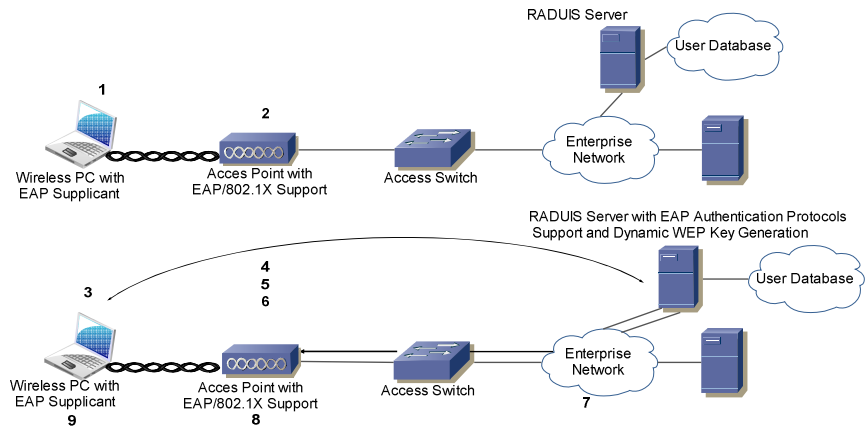

Figure 2. EAP Authentication Process

Here, user login credentials must be provided through an EAP supplicant before wireless client can associate with access point. When association is established, the client and the RADIUS mutually authenticate each other via the access point. The RADIUS server and the client then determine a session key (WEP key) that it distinct to the client. The RADIUS server sends the a session key to the access point via the wired LAN. The access point in turn, encrypts its broadcast key with the session key and sends the encrypted key to the client. The client uses the session key to decrypt it. These keys are valid for all communications during the remainder of the session or until the session expires. A new WEP key is then generated. The session key and broadcast key are changed at regular intervals and can be configured on the access point. Different types of EAP that can be used in wireless network for user authentication include LEAP_Cisco Wireles (LEAP), EAP-Transport Layer Security (EAP-TLS), Protected EAP (PEAP), EAP-tunneled TLS (EAP-TTLS) and EAP-Subscriber.

### 3.4. IEEE 802.11i

IEEE 802.11i defines Robust Security Network (RSN)" [17] that is aimed solving the problems in 802.11b and WEP which include Poor Privacy, lack of encryption key management, Weak authentication and authorization and no Accounting. The RSN adopts an approach that grants the authenticated entity a set of privileges for a limited amount of time. Devices joining in a RSN need to support Counter Mode-CBC MAC protocol (CCMP), a protocol built around the IEEE802.1X standard for access control and AES. These two protocols made RSN a stronger and scalable solution [17]. WEP users using RC4 requires hardware upgrades to join CCMP users. To address this problem RSN allows the use of Temporal Key Integrity Protocol (TKIP), which allows WEP systems to be upgraded to be secure. To support the transition from WEP to TKIP, a network model called transition security network was defined. Transition security Network allows pre-robust security Network associations. This means that WEP users using RC4 can coexist within the same wireless local area network with CCMP users who use AES and TKIP users who use RC4 and security enhancements [18].

### 3.5. Temporal Key Integrity Protocol (TKIP)

TKIP was temporally provided to mitigate the security challenge of WEP in combination with 802.1x authentication and EAP-TLS. It uses 802.IX for key management and establishment. TKIP uses a key scheme based on RC4, but extends this key hierarchy to include a key hash function, and a message integrity check (MIC). The key hash function reduces the exposure of the master secret and to provide per-packet key mixing. TKIP hashes the combination of the IV value, the data encryption key (derived from the master secret), and the MAC address to form the traffic key. This mechanism addresses the WEP problem and then reduces the ability of the related key attack.

### 3.6. Wi-Fi Protected Access (WPA)

Wi-Fi Protected Access is a WLAN data encryption method that uses TKIP to alleviate WEP key flaw by generating a new 128-bit per packet transmitted. [3]. WPA enhances WEP by adding a re-keying mechanism to provide a fresh encryption and integrity key. Temporal keys are changed for every 10,000 packets. This makes it much harder to crack TKIP keys than with WEP. In WPA, a temporal encryption key, transmit address and TKIP Sequence Counter (TSC) form the input to the RC4 algorithm that generates a keystream. MAC Service Data Unit (MSDU) and Message Integrity Check (Michael) are combined using the Michael algorithm. The combination of the MSDU and the MIC is fragmented to generate MAC Protocol Data Units, which is MPDU. From the MPDU, a 32-bit Integrity Check Value is calculated for the MPDU. The combination PDU and the ICV is bit-wise exclusive ORed with a keystream to produce the encrypted data. The IV is added to the encrypted data to generate the MAC frame. According to [19], Michael is not a very strong algorithm. However, it was the best choice given the constraints.

### 3.7. WPA2

This is an enhancement to WPA. It uses AES algorithm for encryption which is stronger than TKIP. AES in combination with Counter Mode with Cipher Block Chaining Message Authentication Code Protocol (CCMP) provide high level security to WLAN. The CCMP algorithm creates message integrity code (MIC) to protect data integrity. WPA2 supports both Enterprise mode and Personal mode. WPA2 Personal uses a set of password. WPA2 Enterprise uses EAP and a RADIUS server for centralized client authentication using multiple authentication methods such as token cards, Kerberos, and certificates.

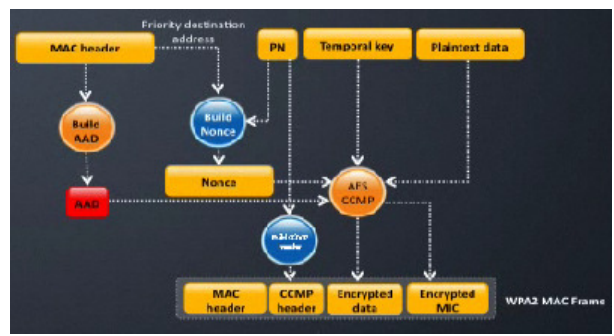

Figure 3. WPA2 CCMP Procedure

During the CCMP procedure, the MAC header produces additional authentication data (AAD) that is added to the CCM encryption process. This protects the frame against alteration of the non encrypted portions of the frame. A sequence packet number PN is included in the CCMP header

35

International Journal of Network Security & Its Applications (IJNSA), Vol.6, No.4, July 2014

to protect against replay attacks. The PN and the portions of the MAC header are used to generate a nonce that is turned, is used by the CCM encryption process. In WPA and WPA2 keys are derived during the four-way handshake shown in Fig. 4 Encryption keys are derived from the Pairwise Master Key. The PMK is derived during the EAP Authentication Session. In the EAP success message, the Pairwise Master Key is sent to the access point, but it is not directed to the Wi-Fi client as it has derived its own copy of the Pairwise Master Key.

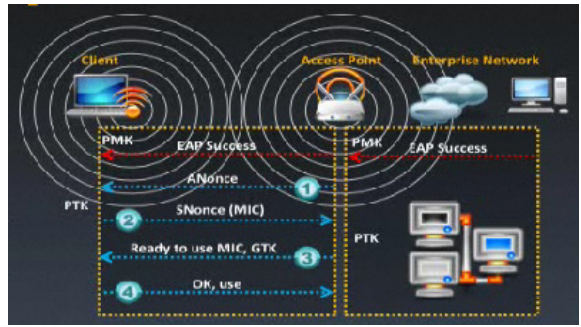

Figure 4. The Four-Way Handshake

1. A client uses the nonce sent by the access point to generate the Pairwise Transient Key (PTK).
2. The client responds with its own nonce value, an Snonce, to the access point together with a message integrity check code which is Michael.
3. The access point sends the groupwise transient key (GTK) and a sequence number together with another MIC which is used in the next broadcast frames.
4. The client confirms that the temporal keys are installed.

## 4. DISCUSSION

Table 1 presented different security standards and solutions such as 802.1x, 802.11i, VPN based SSL and VPN based IPSec, as enhancement to WEP which has several flaws that make the systems that implemented it vulnerable to attackers.

Table 1. Enhanced Wlan Security Comparison

| FEATURES | SSL based VPN | IPsec based VPN | 802.1x EAP with TKIP | | | 802.11i | |
| --- | --- | --- | --- | --- | --- | --- | --- |
| | | | LEAP | PEAP | TLS | WPA | WPA2 |
| Encryption algorithm | RSA key exchange | 3DES or AES | RC4 | RC4 | RC4 | RC4-TKIP | AES-CCMP |
| Key length | 1024- 4096 bit | 168/128, 192, 256 bit | 128 bit | 128 bit | 128 bit | 128 bit | 128-192-256 bit |
| IV length | | | | | | 48 bit | 48 bit |
| Authentication | Mutual | Mutual | Mutual | Mutual | Mutual | Mutual | Mutual |
| Certificate requirements | SSL-enabled Server | Optional | NONE | RADIUS server | RADIUS server/ WLAN client | RADIUS serve | RADIUS serve |
| Packet Integrity | MAC | MD5-HMAC/ SHA-HMAC | CRC-32/MIC | CRC-32/MIC | CRC-32/MIC | MIC | CCM-MIC |
| Key type | Dynamic | Dynamic | Dynamic | Dynamic | Dynamic | Dynamic | Dynamic |
| Key distribution | | Digital certificate | Dynamic | Dynamic | Dynamic | Dynamic | Dynamic |

| Single sign-on support | Yes | Yes | Yes | No | Yes | Yes | Yes |
|---|---|---|---|---|---|---|---|
| Open standard | NO | Yes | No | IETF draft RFC | Yes | IEEE 802.11i | IEEE 802.11i |
| **ATTACKS** | | | | | | | |
| MITM (active attacks) | Alleviated | Vulnerable | Alleviated | Alleviated | Alleviated | Vulnerable | Alleviated |
| Dos | Vulnerable | Vulnerable | Vulnerable | Vulnerable | Vulnerable | Vulnerable | Alleviated |
| Rogue access points | Alleviated | | Alleviated | Alleviated | Alleviated | Vulnerable | Alleviated |
| Passive attacks (FMS paper) | Alleviated | Alleviated | Alleviated | Alleviated | Alleviated | Vulnerable | Alleviated |
| Brute-force dictionary attacks | Vulnerable | Alleviated | Vulnerable with weak password | Alleviated | Alleviated | Vulnerable with weak password | Alleviated |
| Authentication forging | Alleviated | Alleviated | Alleviated | Alleviated | Alleviated | Vulnerable with weak password | Alleviated |
| **OSI LAYER** | Layer 4 | Layer 3 | Layer 3 | Layer 3 | Layer 4 | Layer 2 | Layer 2 |

These enhanced solutions offer security at different layers of Open System Interconnect (OSI) model and provide confidentiality**,** integrity, authenticity and availability which is the basic security requirements needed in a secured network. Each uses encryption algorithm and authentication algorithm shown in the table to ensure secured network. The performance of any of the solutions is dependent on the algorithm and protocols supported. Key length supported by each of these WLAN solutions are shown in the table. Authentication between client and server is mutual but in SSL Client authentication is optional. Using a password-based scheme should require the use of some form of mutual authentication so that the authentication process is protected against brute-force dictionary attacks. Brute-force dictionary attacks against LEAP can be mitigated if strong password is used. It is a good practise for IT administrators to reduce the number of login attempts before an account is locked. Users should be educated on the danger of passing out their username and password to any screen that pop up. The use of one time token should be adopted in build application that run online. When implementing EAP, dictionary attacks can be avoided by using non-password-based schemes such as biometrics, certificates, OTP, Smart Cards, and token cards.

According to [12], WPA2 is vulnerable to RF jamming attack, Layer 2 session hijacking and data flooding attack. In addition, management frames report network topology and modify client behaviour. This facilitates the ability of an attacker to see network layout, discover device location and launch Denial of Service (DoS) attacks. Various researches reviewed in section 2 of this work show that the enhanced solutions and standards are still vulnerable to various attacks they are meant to fight but has made hacking process difficult for the attacker. Unless administrators protect their wireless infrastructure with proven tools and techniques, engage in security awareness training, establish standards and policies that identify proper deployment and security methodology, the integrity of wireless networks will be threatened.

The nature of Wireless communication (over the air) made it vulnerable to passive, active, MITM ,and jamming attacks. A passive attack occurs when someone eavesdrops on network traffic. Cheap availability of wireless network adapter that supports promiscuous mode as well as easily available tools such as Network Monitor in Microsoft products, TCPDump in Linux-based products, NetStumbler, or AirSnort which can be used to capture network traffic for analysis make detecting and reporting on wireless networks a hobby for many wireless wardriving enthusiasts. Disabling SSID broadcast will make the AP not respond to "empty set"SSID beacons and will consequently be "invisible" to network traffic analysis tools. However, it is still possible to capture the "raw"802.11b frames and decode them using programs such as Wireshark (formerly Ethereal) and Wild Packet's AiroPeek to determine the information. Once an attacker has gained sufficient information from a passive attack, they can launch an active such as spoofing, DoS, flooding attacks, introduction of malware and the theft of devices against the network. DoS and flooding attacks can take place once an attacker has authenticated and associated with a wireless network. This can be achieved through the placement and use of rogue access points that prevent wireless traffic from being forwarded properly. Placing a rogue AP within range of a wireless station is a wireless-specific variation of a MITM attack. This attack can be used to gain valuable information about a wireless network, such as authentication requests, the secret key, read private data from a session or to modify the packets thus violating the integrity of a session. Because of their undetectable nature, the only defense against rogue APs is vigilance through frequent site surveys (using tools such as directional or parabolic dish antennae, GPS receivers NetStumbler, Wireshark. and AiroPeek,) and physical security. Rogue access point detection provide a first-level of defence against someone exploiting an access point, but even if they do exploit it, deploying SSL VPN will provide additional layer of security through protection of critical applications and data. It should be noted that NetStumbler will not identify other DoS attacks or other non-networking equipment that is causing conflicts (such as wireless telephones, wireless security cameras, amateur TV (ATV) systems, RF-based remote controls, wireless headsets, microphones and audio speakers, and other devices that use the 2.4 GHz frequency). Spoofing and unauthorized attacks can be protected using MAC filtering to allow only clients that possess valid MAC addresses to access the wireless network. However, MAC addresses are sent in the plain on wireless networks and therefore can easily be changed using edit of the registry in windows or a root shell command in UNIX. A better protection requires the following additional measures; using RADIUS or SecurID, use of VPN to access the wired network, allow only SSH access or SSL-encrypted traffic into the network and use of firewall to isolate unwanted access.

### 4.1. Enhanced 801.11 Security Protocols Performance Evaluation

Strength of algorithm used in encryption, authentication and integrity was used to measure the performance of the enhanced 801.11 Security Protocols. They are weighted on a four point scale (weak, strong, fairly strong and very strong). In addition, susceptibility to attacks and cost of implementation are also indicated.

Table 2. Performance Evaluation for Enhanced 801.11 Security Protocols

| METRICS | SSL based VPN | IPsec based VPN | 802.1x EAP with TKIP | | | 802.11i WPA | WPA2 |
|---|---|---|---|---|---|---|---|
| Algorithm Strength | | | LEAP | PEAP | TLS | | |
| Encryption | **Very Strong** (RSA) | **Very Strong** (3DESor AES) | **Weak** (RC4) | **Weak** (RC4) | **Weak** (RC4) | **Strong** (RC4-TKIP) | **Very Strong** (AES-CCMP) |
| Authentication | **Strong** (Digital Signature) | **Strong** (AH &ESP) | **Weak** (password) | **Fairly Strong** (PEAP) | **Strong** (Digital Certificates) | **Strong** (EAP) | Strong (EAP) |
| Packet Integrity | **Strong** (MAC) | **Strong** (MD5-HMAC/ SHA-HMAC) | **Fairly Strong** (CRC-32/ MIC) | **Fairly Strong** (CRC-32 /MIC) | **Fairly Strong** (CRC-32 / MIC) | **Fairly Strong** (MIC) | **Very Strong** (CCM-MIC) |
| **Susceptibility to Attacks** | Low | Low | High | Low | Low | Low | **Very Low** |
| **Cost of Imple - mentation** | High | High | Low | Low | Moderate | Low | High |

1. **Strength of Algorithm**

VPN ensures a higher degree of confidentiality for traffics. It is difficult for hacker to crack the VPN encryption to see the corporate traffics. If a wireless device is stolen and the theft unreported, the thief would have to know the user credentials to gain access to the VPN. IPSec and SSL are commonly used protocols in VPN to provides for confidentiality of IP traffic, and secure data during transmission . IPSec achieves confidentiality using 3DESor AES. SSL secure most sensitive online transactions such as online banking. It uses digital certificate for authentication, RSA algorithm for encryption and MAC for data integrity

Unlike WEP that that uses pre-shared key for authentication which is easily cracked by attacker, 802.1X uses EAP that provides three significant benefits over WEP, the mutual authentication scheme effectively eliminate MITM attacks. The dynamic distribution of keys prevent lost wireless device from gaining unauthorized access. The centralized policy control enables re-authentication and new key generation. access can us different types of EAP to authenticate users. EAP is a standard authentication method for both WPA and WPA2. There are many types of EAP implementation (LEAP, PEAP, and EAP-TLS) which contributes to its flexibility. LEAP uses passwords for authentication, and does not use digital certificates. Information transported by LEAP is often visible to attackers. User credentials are not strongly protected in LEAP, and are thus easily compromised. PEAP was created as a more secure means of authentication than EAP. It uses an encrypted and authenticated TLS tunnel to send EAP authenticated data. The EAP-TLS algorithm can derive dynamic WEP keys, and the authentication server will send the client the WEP key for use during that session. EAP-TLS is highly secure because it uses certificate-based algorithms and it is hard to crack certificate digitally signed by a CA.

WPA uses RC4 in addition to TKIP. It uses TKIP to encrypt data which enables keys rotation so that every data packet uses a unique encryption key. This makes transmitted data more difficult to hack. TKIP provides a solution to WEP's checksum issues, by using a larger, more secure 48-bit IV, and transmits the IV as an encrypted hash. WPA2 uses AES algorithm for encryption and CCMP for data encryption. CCMP is currently the strongest protocol for encrypting data in a WPA2 network. AES provides much stronger encryption than RC4 and TKIP.

2. **Susceptibility to Attacks**

In 802.11X**, c**ertain "flavors"of RADIUS servers and Web servers can be compromised by buffer-overflow attacks. A buffer-overflow attack occurs when a buffer is flooded with more information than it can hold. The extra data overflows  into other buffers, which may be accessible to hackers. Thus, IEEE 802.1x is susceptible to several attacks, due to the following vulnerabilities:

1. The lack of the requirement of strong mutual authentication. While EAP-TLS does provide strong mutual authentication it is not required and can be overridden.
2.  The vulnerability of the EAP Success message to a MITM attack.
3. The lack of integrity protection for 802.1x management frames.

However, the networks are not as vulnerable as they would be without EAP and 802.1x. Vulnerabilities seen in other solution in Table 2 are discussed in section 2 of this paper. They are comparatively low when compared with WEP. WPA2 still remains the most current secured layer 2 protocol .

3. **Cost of Implementation**

Solutions based on 802.11X are generally more cost effective than those based on VPN for two reasons: First, with 802.1X, the infrastructure is already put in place is augmented. Secondly, the complexity associated with setting up and maintaining VPNs may require more of staff time to manage. Distributing VPN access to remote sites requires installation or adding capacity to a VPN server at that site, and setting up all WLAN users outside the firewall. It is easy to distribute 802.1X-based WLAN access to separate departments, floors and other off-site locations with little administrative overhead.

In WPA, transition from WEP to TKIP is cheap when compared to WPA2. The 802.11i standard requires AES as a replacement for the compromised RC4 algorithm. However, because of the additional processing power required for AES encryption, the addition of a co-processor is likely necessary in wireless device hardware and will make upgrading legacy wireless devices to support AES very costly. As with all other security measures, administrators and managers will have to compare the costs of implementation against the threats the implementation will mitigate.

### 4.2. Using a Defence-in-Depth Strategy

Agencies can mitigate risks to their WLANs by applying Defence-in-Depth Strategy as shown in Fig.5 to address specific threats and vulnerabilities.

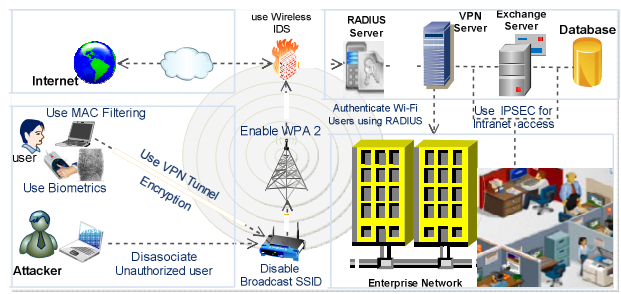

Figure 5. Security In-Dept Approach

The defense-in-depth strategy specifies the use of multiple layers of network security. This can be achieved by

1.  Collecting, analyzing and assessing security events in real time using Wireless Intrusion Detection System (IDS) and a security information and event management (SIEM) solution such as HP's ArcSight. Additional layer of security can be applied the network by disabling Dynamic Host Configuration Protocol ( DHCP).
2.  Per-packet authentication using RADIUS and centralized encryption and security management using SSL VPN for connection security .
3.  WPA2 and AES for Data Protection. Also Using HTTPS instead of HTTP will ensure application layer encryption for a session with a secure web server
4.  use vulnerabilities and patches for device security
5.  deploy a stateful per user firewall for End user protection
6.  auditing, and VPN, IPSec, biometrics for access control. Use strong authentication for network access protection .
7.  Use of encrypted proxy services that allows access to the web server, but with an encrypted tunnel at the local end.
8.  Change all default settings, such as passwords and service set identifiers, or SSIDs.

Biometrics provides added layer of protection and can be integrated with wireless smart cards or other wireless device. It can as well combine with VPN solutions to provide authentication and data confidentiality. It is very difficult to guess physical human characteristics and this made

40

biometric-based authentication highly reliable when compared to password-based authentication. Other counter measures include disable the network when not required, and place wireless access points in a secured location. Planning the Wi-Fi RF coverage area include proper placement of wireless access points and appropriate shielding using thin layer of aluminum under the drywall within the building where possible. Control all access points' power levels, but not to level that legitimate device on the network can no longer connect. These will substantially alleviate RF jamming attack and makes the network highly available. Some wireless interface cards or their drivers cannot capture any packets that are not addressed to the device. Such devices would not allow sniffing on the wireless portion of the network.

It is necessary to highlight that Wi-Fi security standards cannot handle AP failure. They equally lack capacity to handle attacks on physical layer since they provide security on layer two and above.

## 5. CONCLUSIONS

This paper evaluated alternative security mechanisms for WLAN; 802.1X and 802.11i , VPN, IPSec, and SSL. The result of the discussion shows that there is no one solution that is the best. Security problem and mitigation is a continuous process as long as swift changes in wireless technology exist. However, WPA2 is still far more secure than other 802.11options and is still in common use. Therefore achieving secured WLAN requires implementing security at all layers which include wireless signal security, connection security, data protection, device security, network protection, and end user protection. Organizations should perform a risk analysis of their network, develop, and implement relevant and comprehensive security policies throughout their network. Users should be educated on the operation, security and safe computing practices of both wired and wireless networks. This work recommended the use of biometric cryptosystem in place of user name and password for a more secured user authentication. Quantum cryptography is also a likely solution for future machines with exponentially more processing power than today's technology.

**Author**

**Ifeyinwa A. Ajah** is a lecturer in the Computer Science department of Ebonyi State University, Abakaliki Nigeria. She has B.Sc. in Computer Science, M.Sc in Computer System Engineering and Ph.D. in Computer Science. Her research interests include networking, software engineering, internet programming, and database. Dr. Ajah was a recipient of Ambassador for ACM in 2014, Award Certificate from Student Associate Scheme University of East London, United Kingdom, in 2005 and a Commendation Letter from the Cross River State NYSC for Outstanding Performance in the Service Year in February 2000.

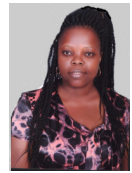